\documentclass[conference]{IEEEtran}

\usepackage[T1]{fontenc}
\usepackage[scaled=0.85]{inconsolata}
\usepackage{microtype}
\usepackage{booktabs}
\usepackage{tabularx}
\usepackage{amsmath}
\usepackage{siunitx}
\usepackage{url}

\usepackage{graphicx}
\usepackage{tikz}
\usetikzlibrary{arrows.meta,positioning}
\usepackage{balance}
\usepackage[hidelinks]{hyperref}

\DeclareSIUnit\celsius{\ensuremath{^\circ\mathrm{C}}}

\title{Toward Lightweight Aerial 5G gNBs: Reproducible OAI Testbed on Commodity ARM Platforms}

\author{%
\IEEEauthorblockN{Marc Duboc and Ammar El Falou}
\IEEEauthorblockA{King Abdullah University of Science and Technology (KAUST)\\
Thuwal, Saudi Arabia\\
Email: \{marc.duboc, ammar.falou\}@kaust.edu.sa}}

\begin{document}
\maketitle

\begin{abstract}
For a battery-powered UAV base station, every gram and watt devoted to RAN-compute
competes with flight-duration margin. We therefore keep the 5G Core (5GC) and
central unit (CU) on the ground and design the airborne unit around only the
backhaul endpoint, distributed unit (DU), and radio. The key question is then a
practical one: can widely available commodity ARM computers sustain a real radio
OpenAirInterface (OAI) 5G DU with useful performance over heterogeneous F1?
We answer it with Raspberry Pi~5 and Jetson Orin Nano as DUs, a USRP B210, a
commercial handset, and Ethernet, Wi-Fi/GRE, and 5G/WireGuard backhaul. Jetson preserves 88.4\% of x86 split-DU Ethernet DL throughput
and 87.5--89.0\% across all three bearers; 5G/WireGuard preserves
76.4--77.1\% of each host's wired DL rate. The validated
Jetson/B210/RM500Q-GL electronics weigh 657.4~g and draw about 28~W under
sustained traffic (757.4~g with integration allowance). A controlled
link-adaptation intervention raises split DL from 23.4 to
99.4~Mb/s as the dominant Modulation and Coding Scheme (MCS) moves from 3 to 26. Finally, synchronized
radio, F1-U, CPU, and UHD evidence narrows the remaining monolithic--split gap
to split-path scheduling/timing behavior. Beyond performance, this capability serves the emergency-response use case
that motivates the aerial cell: once deployed above an affected area, it can
broadcast a PWS warning message. We release the OAI patch that carries the
single-segment Write-Replace Warning over F1 from CU to DU, where SIB8 is
scheduled to the handset. The public artifact makes the
payload, performance, and emergency-broadcast baseline reproducible with
laboratory-accessible hardware.
\end{abstract}

\begin{IEEEkeywords}
5G NR, OpenAirInterface, CU/DU split, F1, AArch64, ARM, testbed,
software-defined radio, wireless backhaul, UAV, aerial RAN, public warning system
\end{IEEEkeywords}

\section{Introduction}
UAV-mounted cellular base stations are attractive for temporary coverage, emergency
response, and rapidly deployable networks, but they face a hard systems
constraint: payload mass and electrical load directly consume the aircraft's
limited energy budget. Carrying a complete base station, therefore spends
valuable airborne resources on functions that do not need to be close to the
radio. The 3GPP F1 split provides a natural boundary: RRC/PDCP/SDAP remain in
the CU on the ground, while the latency-sensitive RLC/MAC/PHY stay in the DU
next to the radio; F1-C uses F1AP/SCTP, and F1-U uses GTP-U/UDP
\cite{3gpp38401,3gpp38473}. The intended airborne payload can consequently be
reduced to a backhaul endpoint, DU, SDR, and antennas rather than an entire
gNB stack. In a disaster area, a particularly valuable first service from
such a temporary cell is an emergency broadcast: PWS can deliver the same
warning or instruction to compatible handsets in the covered area. This makes PWS a natural service requirement for the split aerial architecture.

This architecture only helps if the airborne DU is small and practical
without sacrificing too much radio performance. We deliberately evaluate two
off-the-shelf ARM platforms---Raspberry Pi~5 and Jetson Orin Nano---rather
than custom flight computers. They are widely available, well documented, and
easy for academic laboratories to source, while spanning a useful range from
a low-complexity single-board computer to a higher-performance embedded
platform. The objective is not to argue that ARM is intrinsically superior to
x86; it is to quantify the performance price paid when choosing commodity
embedded compute to make a reproducible airborne DU practical.

This leads to three experimental questions. \emph{RQ1:} how much throughput is
lost when the DU moves from x86 to commodity ARM under the same real-radio
access configuration? \emph{RQ2:} how much additional penalty appears when F1
moves from wired Ethernet to untethered Wi-Fi/GRE or 5G/WireGuard? \emph{RQ3:}
Which lower-layer failures can destroy useful throughput even while F1,
registration, and IP connectivity remain healthy? We answer them with a fixed
B210/UE/n78 access profile, repeated measurements, and a layered acceptance
contract.

The paper answers these questions with three measured results and contributes one additional split-RAN capability:
\begin{itemize}
  \item a reproducible UAV-oriented remote-cell design that keeps 5GC/CU on
  the ground, uses laboratory-accessible ARM as DUs, and measures the integrated
  Jetson/B210/5G-backhaul payload at 657.4~g and approximately consumes 28~W;
  \item a 400-observation host--bearer benchmark that quantifies the throughput cost of embedded compute and heterogeneous F1 under a fixed 106-PRB
  real-radio configuration;
  \item an instrumented failure study: a controlled BLER/MCS intervention
  recovers a 4.25$\times$ throughput collapse, while synchronized mono/split
  telemetry narrows the residual split penalty to scheduling/timing behavior;
  
  \item A documented OAI PWS-over-F1 implementation: a patch that carries the F1AP
  Write-Replace Warning procedure from the ground CU to the remote DU, enabling
  area-wide emergency alerts \cite{pwspatch}.
\end{itemize}
Concurrent two-DU operation is retained as a first-step replication check. The
novelty is therefore a measured and reproducible \emph{airborne
payload/performance trade-off} plus an enabling public-service function for the
split architecture, not the aerial-DU or wireless-F1 concept by itself.

\section{Related Work and Positioning}
Aerial cellular prototypes established the motivation for moving RAN functions
onto aircraft. Flying Rebots, SkyRAN, and SkyCell demonstrated that open SDR
cellular stacks can operate aloft on commodity x86-class compute
\cite{gangula2018,chakraborty2018,ferranti2020}. Their key cost is that substantial base-station compute travels with every
aircraft. An F1 split changes that scaling rule by sharing CU/5GC functions on
the ground and repeating only the remote DU/radio payload.

The closest predecessor is Mundlamuri \emph{et al.}, who demonstrated an OAI
5G aerial DU with a B200mini, Quectel 5G modem, wireless F1, and commercial
terminals \cite{mundlamuri2023}. Our goal is not to show the aerial-DU
concept, but to answer: what happens when the airborne DU is deliberately built from widely reproducible ARM computers, and how much performance is traded for that payload choice across several F1 paths?

Recent non-aerial OAI systems provide complementary evidence. Authors in \cite{bui2025} implement an OAI CU/DU MU-MIMO testbed, while \cite{beurdouche2026} study intensive vRAN deployment across server-class compute architectures. OAI itself provides AArch64 portability
through SIMDe \cite{oaiarmrelease}, and a 2024 6GSPACE report documents
Raspberry Pi execution with the RF simulator \cite{oaiarmpi2024}. These works show that compute matters, but do not provide the same real-radio,
commercial-UE host--backhaul comparison on commodity ARM. Prior PWS work
studies warning-system security \cite{bitsikas2022}. We recently added open-source SIB8 transmission to a monolithic OAI gNB and kept
the implementation modular, specifically to ease future CU/DU support
\cite{abouhasna2026}. We close that disaggregation gap: our published patch implements the F1AP Write-Replace Warning path from CU to DU and validates the resulting SIB8 on a commercial handset \cite{pwspatch}. To the best of our knowledge, no earlier public OAI implementation demonstrates PWS delivery through an actual F1 CU/DU split. Table~I summarizes this section. 

\begin{table*}[t]
\caption{Positioning Against Representative Aerial and OAI Split-RAN Systems. Values across papers are not direct performance rankings because radios, bandwidths, UEs, and traffic differ.}
\label{tab:position}
\centering\scriptsize
\begin{tabularx}{\textwidth}{@{}
>{\raggedright\arraybackslash}p{0.14\textwidth}
>{\raggedright\arraybackslash}p{0.25\textwidth}
>{\raggedright\arraybackslash}p{0.20\textwidth}
>{\raggedright\arraybackslash}X@{}}
\toprule
\textbf{System} & \textbf{RAN / Compute} & \textbf{Experimental focus} & \textbf{Gap Relative to This Work}\\
\midrule
Flying Rebots \cite{gangula2018} & OAI LTE eNB; x86; USRP & Aerial cellular relay & Complete eNB aloft; no 5G F1 split or embedded ARM DU.\\
\addlinespace[2pt]
Mundlamuri \emph{et al.} \cite{mundlamuri2023} & OAI 5G DU; B200mini; Quectel wireless F1 & Aerial DU; $\sim$30 Mb/s to commercial terminals & Closest remote-DU predecessor; no repeated ARM host--bearer matrix or failure diagnosis.\\
\addlinespace[2pt]
Bui and Tran \cite{bui2025} & OAI 5GC/CU/DU; multi-UE NR & MU-MIMO implementation and throughput & Functional CU/DU testbed; not embedded ARM or heterogeneous F1 characterization.\\
\addlinespace[2pt]
Beurdouche and Knopp \cite{beurdouche2026} & OAI vRAN; server-class GPP/accelerator platforms & Intensive multi-instance vRAN compute scaling & Compute-focused central deployment; not remote ARM DUs with B210/UE and variable F1.\\
\addlinespace[2pt]
Abou Hasna \emph{et al.} \cite{abouhasna2026} & Modified monolithic OAI gNB; B210 & Open-source SIB8/PWS security testbed & Public SIB8 broadcast, but CU/DU disaggregation left as future support.\\
\addlinespace[2pt]
This work & Ground 5GC/CU; commodity ARM remote DU; B210; three F1 paths & 400 repeated DL/UL observations; PWS over F1; payload power/mass; failure diagnosis & Reproducible UAV-oriented study plus published CU-to-DU PWS Write-Replace Warning patch.\\
\bottomrule
\end{tabularx}
\end{table*}

\section{Testbed and Method}
\subsection{UAV-Oriented Architecture and Commodity ARM Choice}
\begin{figure*}[t]
\centering
\definecolor{GroundBorder}{HTML}{4B5563}
\definecolor{GroundFill}{HTML}{F9FAFB}
\definecolor{PayloadBorder}{HTML}{1F2937}
\definecolor{PayloadFill}{HTML}{F3F4F6}
\definecolor{RadioBorder}{HTML}{BE123C}
\definecolor{RadioFill}{HTML}{FFE4E6}
\definecolor{RadioText}{HTML}{881337}
\definecolor{FutureBorder}{HTML}{9CA3AF}
\definecolor{Link}{HTML}{1F2937}
\begin{tikzpicture}[x=1cm,y=1cm,font=\scriptsize,
 box/.style={thick,rounded corners=1.5mm,align=center,minimum height=1.55cm},
 link/.style={draw=Link,very thick,-{Stealth[length=2mm,width=1.6mm]}},
 rf/.style={draw=RadioBorder,very thick,-{Stealth[length=2mm,width=1.6mm]}}]
 \path[use as bounding box] (0,0) rectangle (17.9,5.25);

 \draw[thick,rounded corners=2mm,draw=GroundBorder,fill=GroundFill]
   (0.25,0.75) rectangle (5.25,4.35);
 \node[anchor=west,font=\scriptsize\bfseries] at (0.50,4.05)
   {Ground -- shared once};
 \node[box,draw=GroundBorder,fill=white,minimum width=1.75cm] (core) at (1.55,2.55)
   {\bfseries OAI 5GC\\shared core};
 \node[box,draw=GroundBorder,fill=white,minimum width=1.75cm] (cu) at (4.00,2.55)
   {\bfseries OAI CU\\RRC / PDCP\\SDAP};
 \draw[link] (core.east) -- (cu.west);

 \draw[thick,rounded corners=2mm,draw=PayloadBorder,fill=PayloadFill]
   (7.15,0.75) rectangle (14.85,4.35);
 \node[anchor=west,font=\scriptsize\bfseries] at (7.40,4.05)
   {Intended airborne payload -- repeated per cell};
 \node[box,draw=GroundBorder,fill=white,minimum width=1.75cm] (bh) at (8.35,2.55)
   {\bfseries Backhaul\\Eth / Wi-Fi\\RM500Q 5G};
 \node[box,draw=GroundBorder,fill=white,minimum width=1.70cm] (du) at (10.80,2.55)
   {\bfseries ARM DU\\RLC / MAC\\PHY};
 \node[box,draw=RadioBorder,fill=RadioFill,text=RadioText,minimum width=1.90cm] (b210) at (13.55,2.55)
   {\bfseries USRP B210\\n78, 106 PRBs};
 \draw[link] (cu.east) -- node[midway,above,font=\scriptsize\bfseries] {F1-C / F1-U} (bh.west);
 \draw[link] (bh.east) -- (du.west);
 \draw[link] (du.east) -- node[midway,above] {USB 3} (b210.west);

 \node[box,draw=GroundBorder,fill=white,minimum width=1.70cm] (ue) at (16.75,2.55)
   {\bfseries Commercial UE\\data + PWS};
 \draw[rf] (b210.east) -- node[pos=0.65,above,font=\scriptsize\bfseries,text=RadioText] {NR-Uu} (ue.west);

 \node[draw=FutureBorder,dashed,rounded corners=1.5mm,minimum width=4.7cm,
       minimum height=0.55cm,font=\scriptsize\bfseries] (future) at (11.0,4.85)
   {Replicate airborne block: DU$_2$ $\rightarrow$ DU$_N$};
 \draw[draw=FutureBorder,thick,dashed,-{Stealth[length=1.7mm]}]
   (cu.north east) to[out=35,in=180] (future.west);
 \node[font=\scriptsize] at (5.7,0.25)
   {5GC/CU remain on the ground; \hspace{3cm} Only this block is intended to fly.};
\end{tikzpicture}
\caption{UAV-oriented split architecture. Shared 5GC/CU compute remains on the ground, while the repeated airborne unit contains only the F1 backhaul endpoint, commodity ARM DU, and access radio. The solid path is the measured single-DU configuration; concurrent DU$_2$ serves as a supporting check, and larger fan-out is left for future work.}
\label{fig:arch}
\end{figure*}
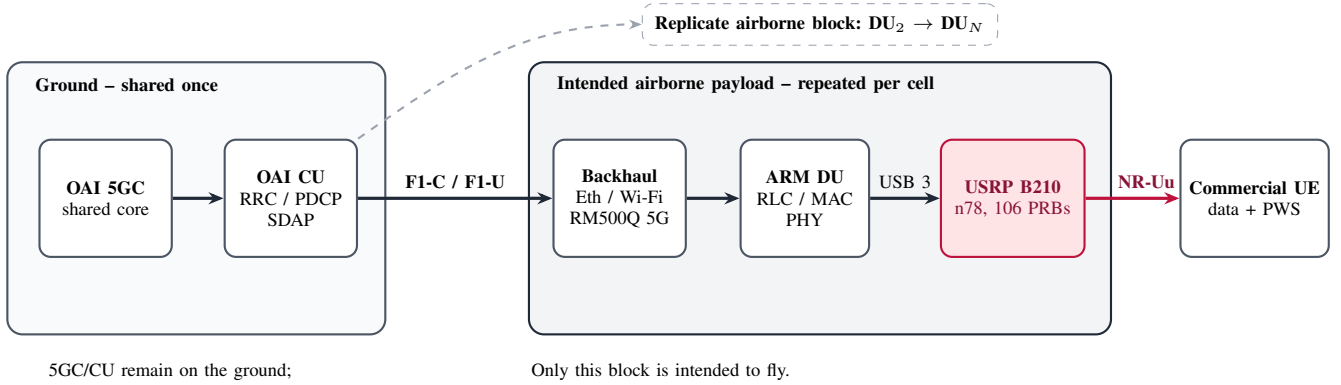

Fig.~\ref{fig:arch} makes the deployment boundary explicit. The x86 ground
host keeps the OAI 5GC/CU; only the backhaul endpoint, ARM DU, B210, and
antennas form the intended airborne unit. The DU runs RLC/MAC/PHY and drives
the access B210 over USB~3. The access cell is fixed at band n78, 3.6192~GHz,
30-kHz SCS, and 106 PRBs, with a commercial Nothing Phone handset as the
end-to-end UE. Ethernet provides the deterministic/tethered reference, while
Wi-Fi/GRE and 5G/WireGuard represent untethered F1 options. In the 5G case, an
RM500Q-GL modem camps on an independent donor gNB, so the access gNB never backhauls itself \cite{artifactrepo}.

\begin{table}[t]
\caption{Experimental Configuration Held Fixed Unless Varied}
\label{tab:config}
\centering\scriptsize
\begin{tabularx}{\columnwidth}{@{}lX@{}}
\toprule
\textbf{Item} & \textbf{Configuration}\\
\midrule
Deployment split & Ground: x86 5GC/CU; intended airborne unit: backhaul + DU + B210\\
OAI & commit \texttt{102965a6}\\
Access radio & USRP B210, USB~3; n78, 3.6192~GHz, 30-kHz SCS, 106 PRBs\\
DU hosts & x86 reference; Raspberry Pi~5 (16 GB); Jetson Orin Nano\\
UE & commercial Nothing Phone handset\\
F1 paths & Ethernet (MTU 9000); Wi-Fi/GRE; RM500Q-GL 5G/WireGuard\\
Jetson SW & L4T R36.4.4 / JetPack 6.2; SCTP-enabled kernel; locked clocks\\
Traffic & iperf3 TCP, one stream, 30 s + 5 s warm-up, separate DL/UL\\
RF control & Faraday cage; same B210/UE/access profile for paired tests\\
\bottomrule
\end{tabularx}
\end{table}

Raspberry Pi~5 and Jetson Orin Nano are chosen intentionally as reproducible
lab platforms rather than bespoke avionics. Pi~5 provides a widely available
low-complexity ARM baseline; Jetson represents the higher-performance embedded
choice. The experiment therefore measures a design trade-off a laboratory can
actually reproduce: how much radio performance is retained when the x86 DU is
replaced by hardware that is realistic to carry and easy to source.

The three F1 modes are implemented as distinct, packet-verifiable paths rather
than aliases over the same management network. The Ethernet profile uses a
9000-byte MTU on the CU/DU path and corresponding core bridge/veth path. The
Wi-Fi profile carries the F1 addresses inside GRE over the wireless underlay.
For 5G/WireGuard, one B210 serves the UE and a second, independent donor
B210 serves the RM500Q-GL; F1 uses the WireGuard inner interface, while packet capture also verifies the modem-facing outer path. Every run checks that F1-C and F1-U do not leak to an unselected Ethernet, Wi-Fi, or management route \cite{artifactrepo}. This matters experimentally because a healthy SCTP association on the wrong interface would otherwise invalidate any backhaul comparison.

The source tree is frozen at commit \texttt{102965a6} and built natively on
ARM through OAI's SIMDe-backed AArch64 path \cite{oaiarmrelease}. Jetson uses
a matching L4T R36.4.4 / JetPack 6.2 SCTP-enabled kernel and rollback entry
\cite{jetsonkernelrepo}. Stable B210 operation requires SuperSpeed USB,
locked clocks, disabled USB autosuspend, bounded DU CPU placement, and xHCI
interrupt isolation; USB~2 produces continuous UHD overflows. These details
are retained because they determine whether the ``lightweight'' platform is a
usable real-radio DU, while the full kernel procedure remains in the artifact.

\subsection{Experimental Control and Measurement}
A run is accepted only if five independent gates pass: (i) core health; (ii)
F1-C/SCTP and F1-U/GTP-U establishment; (iii) packet capture proving that F1
uses the selected bearer rather than a management interface; (iv) UE
registration, PDU session, external reachability, and PWS reception; and (v)
clean teardown and Ethernet rollback. The isolated test PLMN operates inside
a Faraday cage with test-only warning text. Host/radio identity, routes, USB
speed, OAI revision, and patch state are recorded before each run
\cite{artifactrepo}.

Each host--bearer configuration uses 20 repetitions with the same UE, B210,n78/106-PRB profile, OAI revision, and gate sequence. iperf3 TCP uses one
stream for 30 s after a 5 s warm-up; DL and UL are measured separately,
scenario order is randomized, and runs are separated by 10~s. The server is on
the 5GC/CU side, so throughput traverses the complete UE--gNB--5GC user plane.
Reported repeated values are arithmetic means over the 20 valid repetitions; failed service gates invalidate a run rather than create a zero-rate sample.

Absolute RTT is not treated as a bearer metric because it depends on responder placement and ground-network routing. Delay is instead reported as the paired location-induced change $\Delta\mathrm{RTT}(\ell)=\mathrm{RTT}(\ell)-
\mathrm{RTT}(\ell_0)$, where $\ell_0$ is the on-premises CU-side responder and
$\ell$ is the same service at a more distant location; bearer, load, packet size, and test host are held fixed.


\begin{table}[t]
\caption{Measured x86 Split F1-Bearer Throughput Baseline ($n=20$)}
\label{tab:f1baseline}
\centering\scriptsize
\begin{tabularx}{\columnwidth}{@{}Xcc@{}}
\toprule
\textbf{F1 bearer} & \textbf{DL throughput (Mb/s)} & \textbf{UL throughput (Mb/s)}\\
\midrule
Ethernet & $99.4$ & $22.6$\\
Wi-Fi/GRE & $51.8$ & $12.1$\\
5G/WireGuard & $76.2$ & $18.9$\\
\bottomrule
\end{tabularx}
\end{table}

Table~\ref{tab:f1baseline} separates transport context from handset-visible
end-to-end throughput. The untethered paths impose clear throughput penalties,
but delay is not inferred from their raw RTT because that number depends on
endpoint placement and route selection. Delay is therefore reported through location-normalized $\Delta\mathrm{RTT}$,
while bearer substitution remains an end-to-end systems variable including
contention, encapsulation, host networking, and the modem/underlay.

\section{Results}
\subsection{The Cost of Moving the DU from x86 to Commodity ARM}
\begin{figure}
\centering
\definecolor{DLBar}{HTML}{BE123C}
\definecolor{DLBorder}{HTML}{881337}
\definecolor{ULBar}{HTML}{FFE4E6}
\definecolor{ULBorder}{HTML}{BE123C}
\definecolor{GridLine}{HTML}{E5E7EB}
\definecolor{AxisLine}{HTML}{4B5563}
\resizebox{\columnwidth}{!}{%
\begin{tikzpicture}[x=0.032cm,y=0.56cm,font=\scriptsize]
  \foreach \x in {0,50,100,150,200} {
    \draw[GridLine,line width=0.5pt] (\x,-0.60) -- (\x,9.90);
    \node[anchor=north,font=\scriptsize] at (\x,-0.65) {\x};}
  \draw[->,AxisLine,line width=0.7pt] (0,-0.60) -- (206,-0.60);
  \filldraw[draw=DLBorder,fill=DLBar] (20,10.40) rectangle (32,10.75);
  \node[anchor=west,font=\scriptsize\bfseries,text=DLBorder] at (34,10.58) {DL};
  \filldraw[draw=ULBorder,fill=ULBar] (80,10.40) rectangle (92,10.75);
  \node[anchor=west,font=\scriptsize\bfseries,text=DLBorder] at (94,10.58) {UL};
  \foreach \name/\dl/\ul/\row in {
    {mono}/189.2/26.4/9,
    {x86 Eth.}/99.4/22.6/8,
    {x86 Wi-Fi}/51.8/12.1/7,
    {x86 5G}/76.2/18.9/6,
    {Pi Eth.}/61.8/15.4/5,
    {Pi Wi-Fi}/40.9/9.8/4,
    {Pi 5G}/47.2/12.4/3,
    {Jetson Eth.}/87.9/18.1/2,
    {Jetson Wi-Fi}/45.3/13.5/1,
    {Jetson 5G}/67.8/16.6/0} {
      \node[anchor=east,font=\scriptsize] at (-1.5,\row) {\name};
      \filldraw[draw=DLBorder,fill=DLBar] (0,\row+0.05) rectangle (\dl,\row+0.32);
      \filldraw[draw=ULBorder,fill=ULBar] (0,\row-0.32) rectangle (\ul,\row-0.05);
      \node[anchor=west,font=\tiny] at (\dl+1.3,\row+0.18) {\dl};
      \node[anchor=west,font=\tiny] at (\ul+1.3,\row-0.18) {\ul};}
  \node[anchor=north,font=\scriptsize] at (100,-1.35)
    {Handset throughput (Mb/s), arithmetic mean, $n=20$};
\end{tikzpicture}}
\caption{Handset DL/UL throughput across ten configurations (arithmetic mean, $n=20$ per setup; 400 observations).}
\label{fig:throughput}
\end{figure}
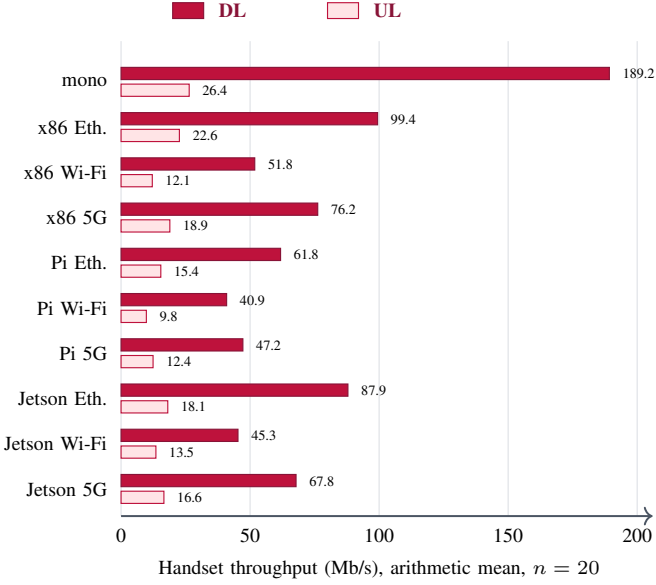

Fig.~\ref{fig:throughput} turns the airborne-compute choice into a measured
trade-off. On Ethernet, x86, Jetson, and Pi~5 split DUs reach 99.4/22.6,
87.9/18.1, and 61.8/15.4~Mb/s DL/UL. Jetson therefore preserves 88.4\% of x86
DL while moving the DU onto an embedded board; Pi~5 preserves 62.2\%. More
importantly, the Jetson/x86 ratio remains 87.5--89.0\% across all three
bearers. Among the tested embedded candidates, Jetson is therefore the
strongest performance-preserving choice for the intended airborne DU, while
Pi~5 provides a simpler reproducible baseline at a larger throughput cost.

The host ranking is the same across Ethernet, Wi-Fi/GRE, and 5G/WireGuard,
showing that the ARM penalty is not a single-bearer accident.

\subsection{The Cost of Untethering F1}
Normalizing each host to its own Ethernet result isolates the transport side
of the design. 5G/WireGuard retains 76.7\% (x86), 77.1\% (Jetson), and
76.4\% (Pi~5) of wired DL; Wi-Fi/GRE retains 52.1\%, 51.5\%, and 66.2\%,
respectively. Thus, the 5G/WireGuard penalty is remarkably consistent across
very different DU hosts in this setup. It should not be read as a general 5G
versus Wi-Fi ranking: the Wi-Fi route is contained, whereas the RM500Q-GL uses an independent donor gNB. Rather, it quantifies the cost of replacing the tethered reference with the actual untethered paths available to this testbed.

UL follows the same broad host ordering but in a narrower dynamic range. This
direction asymmetry matters later: x86 monolithic-to-split DL drops 47.5\%,
whereas UL drops only 14.4\%. Treating ``the split penalty'' as one scalar
would hide that scheduler/traffic-direction effect.

\subsection{A Hidden Failure: Link Adaptation Collapses While F1 Looks Healthy}
In the initial x86 Ethernet split state, F1 and UE attachment remain healthy, but the mean DL is $23.4$~Mb/s. OAI tracks an exponentially filtered BLER,
\begin{equation}
\widehat b_k=0.9\widehat b_{k-1}+0.1b_k,
\end{equation}
and adapts MCS against configured lower/upper bounds \cite{oaimacdoc}. With the
default DL window 0.05--0.15, mean measured BLER is $0.22$, above the upper
bound, and dominant MCS collapses to 3 (2--4).

After observing this state, we choose 0.25--0.35 as a
\emph{diagnostic} DL window that brackets the measured operating BLER; it is
not the result of an optimization sweep and is not proposed as a universal
target. Changing only the BLER thresholds (DL 0.25--0.35, UL 0.15--0.35)
raises dominant MCS to 26 (25--27), mean measured BLER to $0.28$, and mean DL to
$99.4$~Mb/s, a 4.25$\times$ gain. Radio attenuation, BWP, TDD profile,
UE, F1 bearer, and traffic generator are unchanged.

\begin{table}[t]
\caption{Repeated BLER-Threshold Intervention ($n=20$)}
\label{tab:bler}
\centering\scriptsize
\begin{tabularx}{\columnwidth}{@{}Xcc@{}}
\toprule
\textbf{Metric} & \textbf{Default} & \textbf{Diagnostic retune}\\
\midrule
DL target window & 0.05--0.15 & 0.25--0.35\\
DL throughput (Mb/s) & $23.4$ & $99.4$\\
Filtered DL BLER & $0.22$ & $0.28$\\
Dominant MCS (range) & 3 (2--4) & 26 (25--27)\\
\bottomrule
\end{tabularx}
\end{table}

The higher post-retune BLER is therefore not a regression: it lies inside the
relaxed diagnostic window and is paired with a much higher MCS. This
before/after experiment shows why F1 setup and packet flow alone are
insufficient health criteria, while avoiding the stronger claim that
0.25--0.35 is optimal for other channels or UEs.

\subsection{Residual Monolithic--Split Gap}
The x86 monolithic reference reaches mean DL of $189.2$~Mb/s versus
$99.4$~Mb/s for the \emph{retuned} Ethernet split, a 47.5\% reduction;
mean UL falls only 14.4\% ($26.4$ to $22.6$~Mb/s). We compare against
the repaired split state specifically so the default BLER collapse is not
mistaken for the explanation of this residual gap.

\begin{table}[t]
\caption{x86 Monolithic--Ethernet Split Diagnostic Evidence}
\label{tab:monosplit}
\centering\scriptsize
\begin{tabularx}{\columnwidth}{@{}Xcc@{}}
\toprule
\textbf{Metric} & \textbf{Monolithic} & \textbf{Split Ethernet}\\
\midrule
DL throughput (Mb/s) & $189.2$ & $99.4$\\
UL throughput (Mb/s) & $26.4$ & $22.6$\\
Dominant DL MCS & 26 & 26\\
Filtered DL BLER & 0.27 & 0.28\\
F1-U rate at 99.4 Mb/s app. & --- & 104.8 Mb/s\\
F1-U loss / persistent queue & --- & 0 / none\\
Busiest DU RT thread & --- & 62\%\\
UHD O/L/U counters & 0/0/0 & 0/0/0\\
PRB/TB observation & full/continuous & intermittent gaps / smaller TBs\\
\bottomrule
\end{tabularx}
\end{table}

The measured radio operating point is nearly identical (MCS~26; BLER
0.27/0.28). Split F1-U carries 104.8~Mb/s, 5.4\% above application throughput,
with no observed loss or persistent queue; the busiest DU real-time thread is
62\%, and UHD reports no \texttt{O}/\texttt{L}/\texttt{U} events. RF-quality
collapse, Ethernet capacity, persistent F1-U backlog, and sustained gross CPU
saturation are therefore poor primary explanations.

The remaining observed discriminator is scheduling occupancy: monolithic
execution allocates the full 106-PRB budget on each active DL slot, whereas
split execution shows
inter-slot gaps and smaller TBs. The current logs do not retain a repeated
gap-frequency or TB-size distribution, so this evidence is qualitative. It
\emph{narrows the likely cause} to split-path scheduling/polling/timing
behavior but does not distinguish F1 arrival jitter from internal OAI
producer--consumer cadence or establish a software root cause.

\subsection{Airborne Payload, PWS over F1, and Service Validation}
The performance numbers matter because they attach to a concrete payload. For
the validated Jetson configuration, the board, B210 with access antennas, and
RM500Q-GL modem kit total 657.4~g and draw approximately 28~W under sustained
traffic. Adding a 100-g integration allowance gives 757.4~g. We do not convert
28~W into flight time because battery voltage, propulsion load, regulator
efficiency, cooling, and airframe are aircraft-specific; the measurement is
instead the electrical budget that a flight designer must reserve for the
cellular payload.

\begin{table}[t]
\caption{Measured UAV-Oriented Electronics Payload}
\label{tab:payload}
\centering\scriptsize
\begin{tabularx}{\columnwidth}{@{}Xr@{}}
\toprule
\textbf{Item / metric} & \textbf{Measured value}\\
\midrule
Jetson Orin Nano board & 151.4 g\\
USRP B210 + access antennas & 217.3 g\\
RM500Q-GL modem kit + antennas & 288.7 g\\
\midrule
Core electronics mass & 657.4 g\\
Integration allowance (+100 g) & 757.4 g\\
Sustained-traffic electrical load & $\sim$28 W\\
\bottomrule
\end{tabularx}
\end{table}

Public-warning delivery is a distinct software contribution and a direct
emergency-response requirement. A temporary cell deployed above a disaster zone should be able to broadcast safety instructions as soon as coverage is available; PWS is well-suited to this role because the warning is cell-broadcast rather than being individually addressed. In prior monolithic OAI work, SIB8 terminates within a single gNB process. In our split gNB, however, the warning originates at the ground CU while the remote DU owns MAC scheduling and the
radio. Our patch, therefore, adds the missing F1AP Write-Replace Warning path: the CU packages the PWS system-information container and sends the request over SCTP/F1-C; the DU decodes it and hands SIB8 to MAC for SI scheduling \cite{pwspatch}. In the validated single-segment case, the warning traverses CU$\rightarrow$F1$\rightarrow$DU$\rightarrow$SIB8 and is displayed by the commercial handset. The available instrumentation records a 0.13-s CU-to-DU
network-side event and about 0.8~s handset-visible latency; because these
timestamps are not generated from one synchronized clock, we treat them as
service-validation intervals, not a stage-by-stage latency decomposition.

As a secondary replication check, two DUs remain simultaneously active behind
one CU under sustained traffic, retaining 91\% and 93\% of their individual
single-DU baselines. This is a first-step shared-CU feasibility result rather than a scaling study, but together with PWS, it shows that the lightweight remote-cell unit preserves both user data and application-visible broadcast service.

A synchronized Pi~5 sample further uses about 1.8 CPU cores and 1.4~GB RAM at
$64.8\,^{\circ}$C without late/UHD overflow markers after thread pinning. The
B210 remains the validated access radio; a future B200mini-class radio could
reduce payload mass further, but would require a new integrated power/thermal
measurement \cite{ettusb210power}.

\section{Discussion and Design Implications}
The measurements give a direct answer to the UAV design question. Keeping the
5GC/CU on the ground avoids repeating shared compute on every aircraft; the
remaining choice is how much performance to buy with the computer that does
fly. In our fixed 106-PRB testbed, Jetson loses only 10--12.5\% of x86 DL
across the three bearers, while Pi~5 accepts a larger performance reduction.
Combined with the measured 657.4-g / 28-W Jetson-based electronics point, this
makes the Jetson configuration the most attractive \emph{performance-preserving}
embedded option we tested. This is a testbed-specific design conclusion, not a
claim that one ARM board is universally optimal.

Untethering the F1 link creates a second, largely separable cost. The
5G/WireGuard path preserves about 76--77\% of each host's Ethernet DL, whereas
the tested Wi-Fi/GRE route is lower and more variable. For an aerial system,
this means compute and backhaul should be budgeted independently: selecting a
faster DU does not remove the penalty of the wireless F1 path, and a good
backhaul cannot compensate for an underpowered DU.

The BLER experiment adds an operational lesson that matters outside the
headline throughput plot. A cell can have F1 established, a registered UE, and
working IP traffic while link adaptation has collapsed to MCS~3. The
controlled threshold intervention recovers 4.25$\times$ DL without changing
RF, UE, BWP, TDD, or F1. For a remote or airborne cell, such a failure would be
particularly costly because the system would consume flight energy while
providing a fraction of its expected capacity. The layered service gates and
telemetry are therefore part of the deployment design, not merely debugging
convenience.

The remaining x86 monolithic--split gap is separate from that BLER failure.
Both states use MCS~26 with BLER 0.27/0.28; split Ethernet carries
104.8~Mb/s F1-U at 99.4~Mb/s application throughput, with zero observed F1-U
loss/persistent queue, a 62\% busiest DU real-time thread,
and clean UHD counters. These observations make sustained RF, F1-capacity,
queue, or gross CPU saturation poor explanations. The distinguishing trace is
intermittent PRB/TB scheduling in split mode versus continuous full-PRB use
monolithically. We therefore narrow the likely cause to the split-path
scheduling/polling/timing behavior. Quantitative slot-aligned timestamps for
F1-U arrival, DU wake-up, grant construction, TB size, PRB occupancy, and HARQ
feedback are still required to distinguish network arrival jitter from an
internal OAI producer--consumer cadence.

The study intentionally stops before claiming flight endurance or universal
bearer rankings. We test one n78/106-PRB profile, one commercial UE, and
specific Wi-Fi and donor-cell routes. Likewise, the diagnostic BLER window was
chosen after observing the failure state; it proves the controller mechanism
in this setup, not a universal optimum. These boundaries do not weaken the
central result: they define a reproducible laboratory baseline from which an
airframe designer can trade DU performance, backhaul choice, payload mass, and
power without first rebuilding the cellular stack from scratch.

\section{Reproducibility Artifact}
To make the UAV-oriented baseline reproducible outside our lab, the public
artifact pins the OAI revision, the published PWS-over-F1 and instrumentation
patches, sanitized scenario templates, host discovery, MTU/routing setup,
Jetson SCTP kernel build, packet-path checks, service gates, and rollback
workflow \cite{artifactrepo,pwspatch,jetsonkernelrepo}. It also records the operational details
that are brittle in a real lab: live B210 ownership, the RM500Q-GL donor/access
boundary, Ethernet MTU restoration, GRE/WireGuard route selection, and
detection of stale F1 traffic on management interfaces. Reproduction requires
F1 on the intended interface, UE attachment to the access B210, a working PDU
session, and clean return to the Ethernet baseline.

This artifact-first contract is important because several failure modes in the
paper leave processes alive: wrong-interface F1 can still establish SCTP, and the default BLER state still registers the UE and carries IP traffic. The reproduction target is therefore the layered observable signature
(configuration, packet path, radio state, service, and performance), not a
successful launch command. 

\section{Conclusion}
We built the split-DU testbed around a simple aerial design objective: keep
shared 5GC/CU compute on the ground and make the repeated airborne unit as
small and reproducible as possible without giving up useful real-radio
performance. With commodity hardware, Jetson retains 88.4\% of x86 Ethernet
DL and 87.5--89.0\% across all tested F1 bearers, while the validated
Jetson/B210/RM500Q-GL electronics weigh 657.4~g and draw about 28~W. Across
400 repeated measurements, the resulting arithmetic means provide practical
design baselines. A controlled BLER intervention
recovers a 4.25$\times$ hidden throughput collapse, and synchronized
mono/split telemetry narrows the remaining split penalty to scheduling/timing
behavior rather than a simple RF, F1-capacity, or CPU bottleneck. We also
contribute to a public OAI PWS-over-F1 path: a released F1AP Write-Replace Warning patch carries the warning from the
ground CU to the remote DU, which broadcasts SIB8 to compatible handsets
without per-user phone-number addressing. This directly connects the system objective to the emergency use case: a lightweight cell flown over an affected area can provide an immediate area-wide warning service as well as user data.
Two-DU operation further verifies the first-step replication behind the shared ground CU. The result is not only an OAI setup that runs on ARM, but a reproducible remote-cell building block for future UAV experiments where every carried gram and watt matters.

\bibliographystyle{IEEEtran}
\bibliography{research-paper}

\end{document}